\documentclass[showpacs,preprintnumbers,twocolumn,aps,prl,reprint,superscriptaddress,nobibnotes,nofootinbib]{revtex4-1}

\usepackage{amsmath,amssymb,hyperref,times,graphicx,bm,subfigure,bbold,color}
\usepackage{bibunits}
\usepackage{wasysym}
\usepackage{physics} 
\usepackage{calrsfs}
\usepackage{graphicx}

\usepackage{bm}
\usepackage{braket}
\usepackage{color}
\usepackage{bbm}
\usepackage{hyperref}
\usepackage{dsfont}
 
\def\<{\langle}
\def\>{\rangle}

\newcommand{\be}{\begin{equation}}
\newcommand{\ee}{\end{equation}}                  
\newcommand{\bea}{\begin{eqnarray}}
\newcommand{\eea}{\end{eqnarray}}
\newcommand{\beas}{\begin{eqnarray*}}
	\newcommand{\eeas}{\end{eqnarray*}}

\begin{document}

\title{Singularity in Entanglement Negativity Across Finite Temperature Phase Transitions}

\author{Tsung-Cheng Lu} 

\author{Tarun Grover}

\affiliation{Department of Physics, University of California at San
	Diego, La Jolla, CA 92093, USA}
\begin{abstract}
Phase transitions at a finite (i.e. non-zero) temperature are typically dominated by classical correlations, in contrast to zero temperature transitions where quantum mechanics plays an essential role. Therefore, it is natural to ask if there are any signatures of a finite temperature phase transition in measures that are sensitive only to  quantum correlations. Here we study one such measure, namely, entanglement negativity, across finite temperature phase transitions in several exactly solvable Hamiltonians and find that it is a singular function of temperature across the  transition. As an aside, we also calculate the entanglement of formation exactly in a related, interacting model.
\end{abstract}

\maketitle

Interacting quantum systems with competing interactions can exhibit phase transitions at both zero and non-zero temperatures. Heuristically, the zero temperature phase transitions result due to quantum fluctutions while the finite temperature phase transitions typically result from thermal fluctuations\cite{sachdev2011quantum}. As an example, consider the 2+1-D transverse field Ising model on a square lattice: $H = -\sum\limits_{<i,j>}^{} Z_i Z_j - h \sum\limits_{i}^{}X_i$. Here the critical exponents associated with the zero temperature phase transition belong to the three dimensional Ising universality while those for the finite temperature phase transition belong to the two dimensional Ising universality \cite{suzuki1976,sachdev2011quantum}. That is, at any non-zero temperature, the universal critical exponents are identical to those corresponding to the purely classical Hamiltonian $H = - \sum\limits_{<i,j>}^{} Z_i Z_j$. Given this observation, it is natural to ask are there any singular correlations at a finite temperature transition that are intrinsically quantum-mechanical? For a pure state, von Neumann entanglement entropy is a good measure of quantum correlations, but since we are interested in finite temperature transitions, we need to consider measures of mixed state entanglement. With this motivation, in this paper we will introduce certain quantum models which exhibit finite temperature transitions, and we will analytically study mixed state entanglement measures in these models, with a particular focus on entanglement negativity \cite{vidal2002}.

One way to motivate mixed state entanglement measures is via the notion of `separable' states - these are states that can be prepared from any other state using only local operations and classical communications (LOCC), and therefore are not entangled. A bipartite mixed state is separable if it can be written as $\rho = \sum_i p_i\, \rho_{A,i} \otimes \rho_{B,i}$ where $p_i > 0$ while $\rho_{A,i}, \rho_{B,i}$ are valid density matrices \cite{nielsen2002, werner1989}. For \textit{pure} states, the von Neumann  entropy $S = - \tr \left( \rho_A \log(\rho_A)\right)$, where $\rho_A$ is the reduced density matrix on Hilbert space $A$, is a faithful measure of quantum correlations. However, $S$ is rather ineffective at capturing mixed state quantum correlations. For example, even a thermal density matrix corresponding to a purely \textit{classical} Hamiltonian will have a rather substantial von Neumann  entropy $S$ that equals the thermal entropy for region $A$. Several measures of mixed state entanglement have been proposed (see, e.g., Ref.\cite{horodecki_revmodphys} for an overview) including entanglement of formation, entanglement of distillation, entanglement of purification, squashed entanglement and entanglement negativity. As yet, all of these measures, with the  exception of entanglement negativity, require optimizing a function over all possible quantum states, making their calculation rather challenging. Therefore, below we will primarily focus on the entanglement negativity with one exception; for a specific many-body model we will also  calculate  the entanglement of formation.

The entanglement negativity (henceforth, just `negativity' for brevity) is defined as follows \cite{eisert99, vidal2002}: given a bipartite density matrix $\rho$  acting on the Hilbert space $\mathcal{H}_A \otimes \mathcal{H}_B$, one first performs a partial transpose \textit{only}  on the Hilbert space $\mathcal{H}_B$ to obtain a matrix $\rho^{T_B}$. Explicitly, if $\rho = \sum\limits_{A,B,A',B'} \rho_{AB,A'B'} |A\rangle |B\rangle \langle A'| \langle B'| $, then $\rho^{T_B}= \sum\limits_{A,B,A',B'} \rho_{AB,A'B'} |A\rangle |B'\rangle \langle A'| \langle B|.$ The matrix $\rho^{T_B}$ is Hermitian but is not necessarily positive semi-definite. The  negativity $E_N$ is defined as $E_N = \log\left(||\rho^{T_B}||_1\right)$. The utility of this procedure becomes apparent when one notices that negativity is zero for separable mixed states \cite{werner1989, peres1996, horodecki1996, simon2000, vidal2002}. This is because for separable states, $\rho^{T_B}$ is a valid density matrix, and therefore, $||\rho^{T_B}||_1 = 1$. The main drawback of negativity is that it can be zero even for non-separable states \cite{horodecki1997}. Heuristically, this means that although negativity is insensitive to classical correlations, it does not capture \textit{all} quantum  correlations. Since we will also briefly discuss entanglement of formation, denoted as $E_F$, let us also recall its definition. $E_F$ for a bipartite mixed state $\rho_{AB}$ is defined as follows  \cite{bennett1996}:  decomposing $\rho_{AB}$ as a convex sum of pure states, $\rho_{AB} = \sum_i p_i |\psi_i\rangle \langle \psi_i|$ where $p_i > 0$ with $\sum_i p_i = 1$, $E_F$ is given by $E_F = \textrm{inf}\, \sum_i p_i S(\textrm{Tr}_B |\psi_i\rangle \langle \psi_i| )$ where $S$ is the von Neumann  entropy. Therefore, $E_F$ is the least possible entanglement of any ensemble of pure states that realizes a given mixed state. In contrast to negativity, $E_F$ is zero if and only if a state is separable.

To begin with, we note  one feature of  negativity shared by all Hamiltonians considered here, as well as in several other lattice models (see, e.g., Refs.\cite{anders2008, ferraro2008, sherman2016, castelnovo2018}) and continuum field theories \cite{calabrese2012, calabrese2015}: above a certain temperature, the negativity for the corresponding thermal (Gibbs) state becomes exactly zero. This temperature is called `sudden death temperature' denoted as $T_d$. One of the central questions we will ask is the following. Consider an interacting system which exhibits spontaneous symmetry breaking below a critical temperature $T_c$. Assuming that negativity $E_N$ is non-zero in the vicinity of the transition (i.e. the condition $T_d > T_c$ is satisfied), is $E_N$ a singular function of the tuning parameter (e.g. the temperature) across the transition? 

We now state our main result. We find that in all models considered in this paper, whenever negativity is non-zero in the vicinity of the transition, it is always singular across the transition. This result is at variance with expectations from Ref.\cite{sherman2016}  where numerical calculations on finite sized systems for the 2+1-D quantum Ising model suggested that negativity is analytic across the corresponding $T_c$. We will return to a comparison with Ref.\cite{sherman2016} after discussing our results.

\begin{figure}
	\subfigure{\label{fig:a}\includegraphics[width=0.45\textwidth]{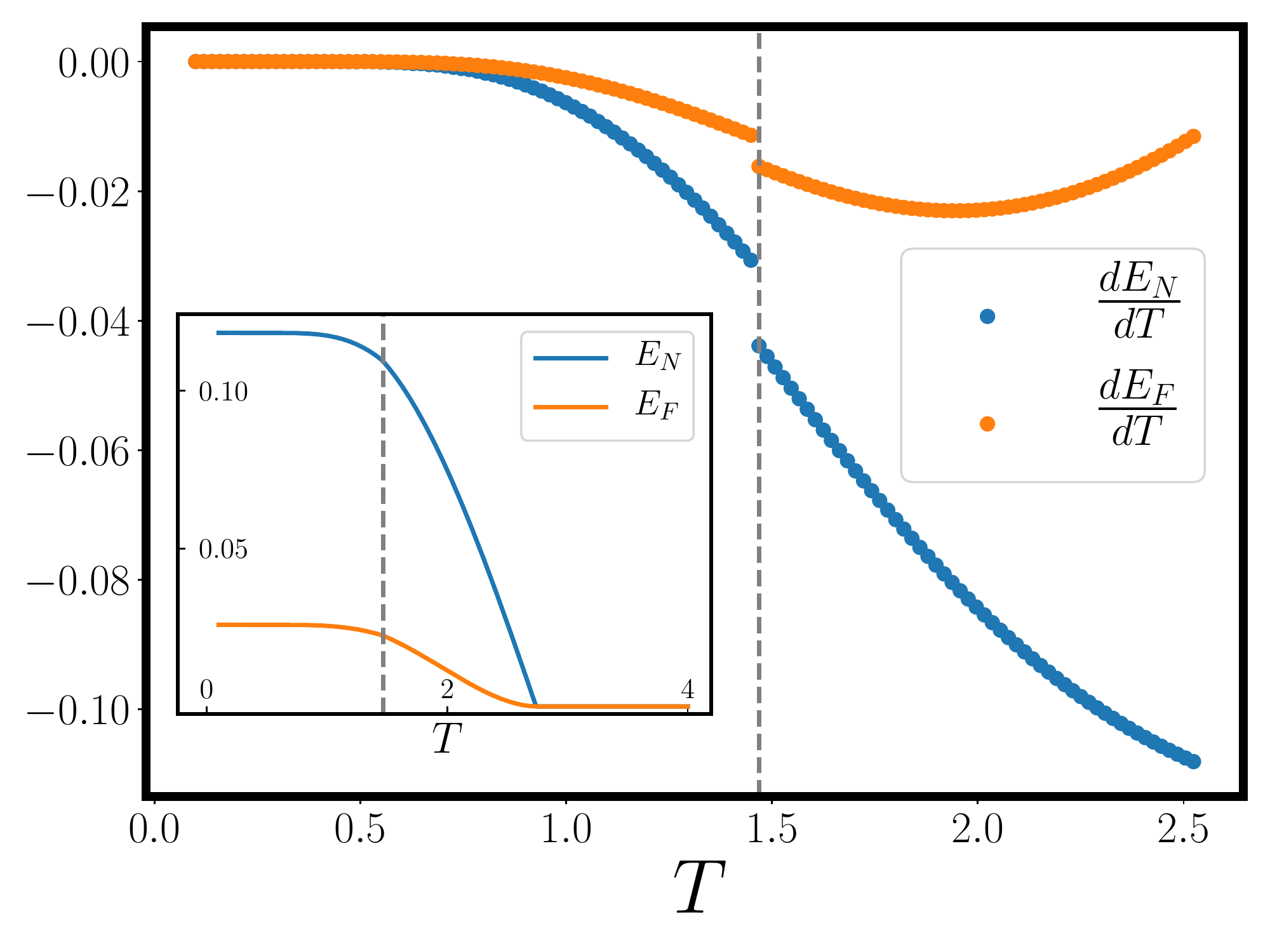}}
	\caption{The derivative of entanglement negativity $\frac{dE_N}{dT}$ and the derivative of entanglement of formation $\frac{E_F}{dT}$ corresponding to the thermal state for a two site mean-field theory of the transverse field Ising model, Eq.\ref{eq:Hmeanfield} with $h=3.8$ and $z=2$. Inset:  $E_N$ and $E_F$ for the two site mean-field theory for the same problem. The vertical dashed gray line in both plots indicates the location of the critical temperature.} \label{fig:meanfield}
\end{figure}

As a starting point, consider a single site mean-field Hamiltonian for the transverse field Ising model: $H^{MF}_{\textrm{1 site}} = -  mz Z - h X$, where $z$ is the coordination number. The corresponding thermal state is indeed separable, which might  lead one to expect that perhaps negativity is always an analytic function across  finite temperature transitions. However,  a single site mean-field is too crude an approximation: within such a mean-field approximation, even the  ground state is unentangled and shows no singularity in the quantum entanglement across a T = 0 quantum phase transition (QPT), in contrast to the known exact results (see, e.g., Refs.\cite{nielsen2002a, metlitski2009, kallin2013}). To improve upon this, we next consider a \textit{two-site} mean-field theory:  

\be 
H^{MF}_{\textrm{2 sites}} = - m(z-1) \left(Z_1 +Z_2\right) - Z_1Z_2 - h \left(X_1 +X_2 \right) \label{eq:Hmeanfield}
\ee  
and study the  negativity for a bipartition that runs across the two sites. A straightforward calculation shows that whenever $T_d > T_c$, the critical temperature for the phase transition, the negativity \textit{is} a singular function of the temperature across the transition, see Fig.\ref{fig:meanfield}. Incidentally, since an analytical expression for entanglement of formation $E_F$ is available for any state acting on two qubits \cite{wooters1997}, we calculate $E_F$ as well for this mean-field model, and find that it is also singular across the transition (Fig.\ref{fig:meanfield}).

 \begin{figure}
	\centering
	\includegraphics[width=0.45\textwidth]{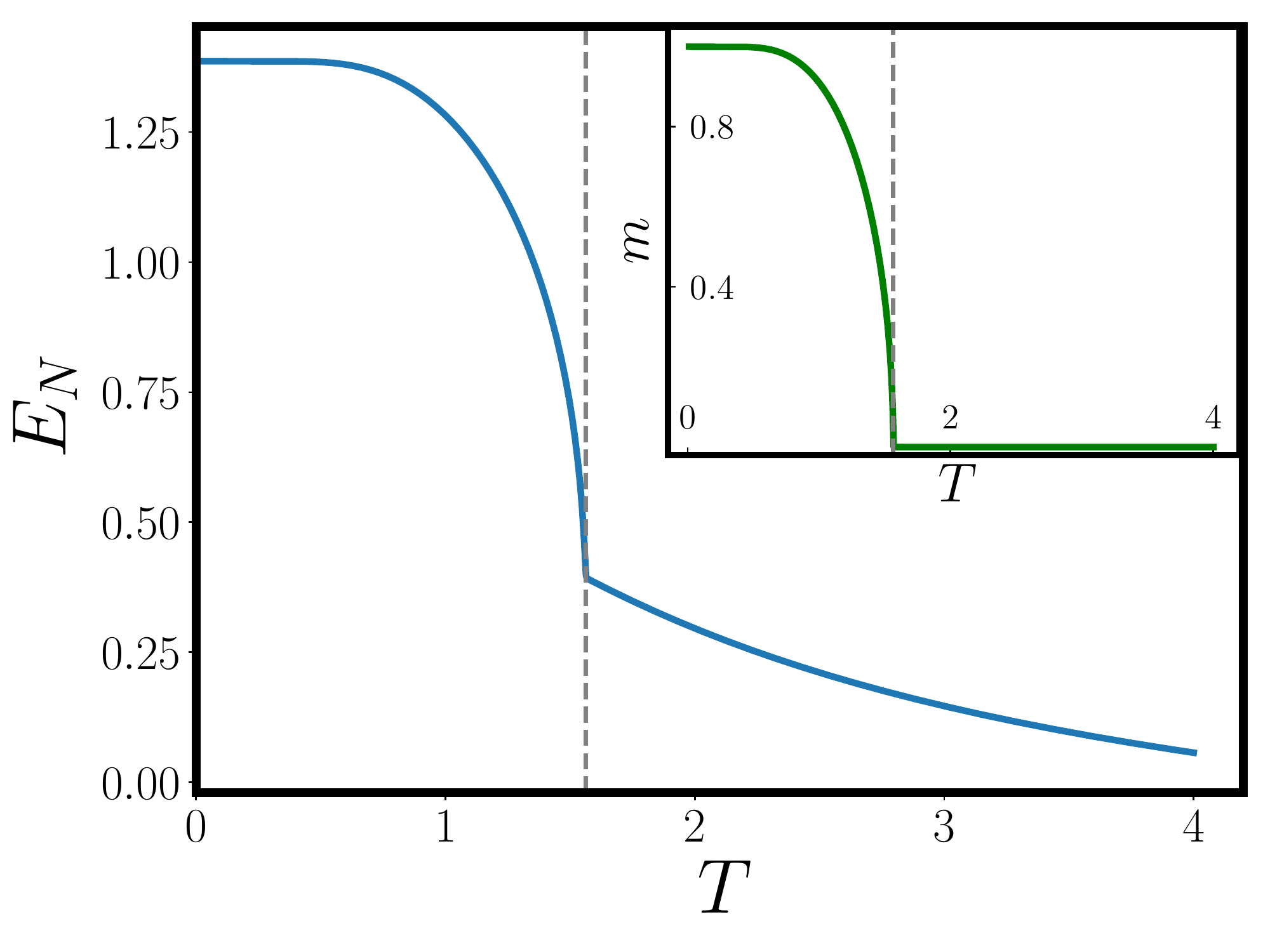}
	\caption{Negativity as a function of temperature for the non-local commuting projector model, Eq.\ref{eq:H_nonlocal_stab}, for $g_x = 2, g_z = 1$. Inset: Temperature dependence of the magnetization. The vertical dashed gray line in both plots indicates the location of the critical temperature.}
	\label{fig:inf_stabilizer}
\end{figure}

Motivated by the two-site mean-field result and the models studied in Ref.\cite{assaad2016}, we next consider a Hamiltonian which exhibits a finite temperature transition, and where negativity is calculable exactly in the thermodynamic limit. The model is defined on a one-dimensional lattice with $L$ sites where each lattice site has four qubits: 
\bea\label{eq:inf_stabilizer}
 H& =-& \frac{1}{4L}  \left( \sum_i  \left( Z_{i1}   Z_{i2} + Z_{i3} Z_{i4}\right)   \right)^2 -g_z \sum_i  Z_{i1}   Z_{i2} Z_{i3} Z_{i4}  \nonumber \\
 & & -g_x \sum_i  \left( X_{i1}   X_{i2}+X_{i3} X_{i4}\right). \label{eq:H_nonlocal_stab}
 \eea 
 The most notable feature of this Hamiltonian is that it is a sum of commuting terms, and it supports a finite temperature transition where the Ising symmetry corresponding to $Z_{i1} Z_{i2} \rightarrow - Z_{i1} Z_{i2}, Z_{i3} Z_{i4} \rightarrow -Z_{i3} Z_{i4}$ gets spontaneously broken. The first term in the Hamiltonian makes it non-local and leads to a finite temperature Ising transition in the mean-field universality class. Defining the order parameter $m = \< Z_{i1}   Z_{i2} \>= \<Z_{i3}   Z_{i4} \>$, one finds that in the thermodynamic limit, the critical temperature is given by the solution of the equation $ 2 \beta  = 1 + e^{- 2 \beta g_z}$ while the order parameter $m$ is determined via $\frac{\sinh(2\beta m)}{  \cosh(2\beta m )   +e^{-2\beta g_z}  } =m$ which implies that close to $T_c$, $m = a \sqrt{T_c -T}$, as expected. 
 Next, we calculate the negativity of this model for the bipartition that runs across the four qubits on a chosen site, i.e., $A = \{i_\alpha, i < 0, \alpha = 1,2,3,4 \} \cup \{i=0, \alpha = 1,3\}$ and $B = \overline{A}$ where we have chosen the cut across the site 0 for convention. One finds that for all $T \geq T_c$, and for $0 \leq (T_c-T)/T_c \ll 1$, the negativity is given by $E_N = \log(1+F)$, where 
 
 \be
 F = \frac{e^{\beta g_z} \sinh(2\beta g_x)\cosh(2\beta m)-e^{-\beta g_z}\cosh\left(2\beta g_x  \right)  }{  2\left(   \cosh(\beta g_x)\right)^2  \left(e^{\beta g_z}\cosh(2\beta m ) +e^{-\beta g_z}    \right) } 
 \ee
 assuming $F > 0$  \cite{supplement}; otherwise negativity is zero which also yields an expression for $T_d$ by setting $F = 0$. Since the critical temperature $T_c$ depends only on $g_z$, one can always tune $g_x$, so that the sudden death temperature is higher than $T_c$. Since $m$ is a singular function of temperature so is negativity. In fact $E_N$ simply inherits the cusp singularity of $m$ across the phase transition, i.e., $\frac{dE_N}{dT}|_{T = T^{-}_c} \neq \frac{dE_N}{dT}|_{T = T^{+}_c}$, see Fig.\ref{fig:inf_stabilizer} which also shows the temperature dependence of negativity for all temperatures including $T < T_c$.
 
One drawback of the model just discussed is that it is non-local and relatedly, exhibits mean-field scaling exponents. Therefore, it would be worthwhile to study negativity in thermal states of local Hamiltonians that host a finite temperature transition. Before considering local models, we notice a property specific to commuting projector models that will simplify our subsequent discussion. Let's decompose a commuting projector Hamiltonian as $H=H_A+H_B+H_{AB}$, and further denote the Hilbert space of spins on the boundary of region $A(B)$ that interact with $B(A)$ by $\partial A (\partial B)$.  We also define $A'= A - \partial A, B' = B - \partial B$ i.e. spins strictly in the `bulk' of $A (B)$. Note that $H_{AB}$ acts only on the boundary Hilbert space of  $\partial A, \partial B$. One can show that \cite{supplement}

\be\label{eq:reduced_nega} 
\norm{\rho^{T_B} }_1 =\norm{\rho_{\partial A,\partial B}^{T_{\partial B}} }_1
\ee 
where $\rho_{\partial A,\partial B} = \frac{1}{Z} \Tr_{A',B'}  e^{-\beta H} = \frac{1}{Z} e^{-\beta H_{AB}}   \Tr_{A',B'}     e^{-\beta \left(H_{A}+H_B \right)  }$ is the reduced density matrix for the boundary spins. This property results from the fact that partial transpose  affects operators only at the boundary (i.e. only in the factor $ e^{-\beta H_{AB}} $ in the expression for $\rho_{\partial A,\partial B}$), and furthermore one can always find a basis in which $H_A$, $H_B$ and  $\left( e^{-\beta H_{AB}}   \right)^{T_{\partial B}}$ can all be simultaneously diagonalized.

With the aforementioned property specific to commuting projector Hamiltonian , we now turn our attention to the negativity in a local Hamiltonian defined on a square lattice, with two species of spins, $a$ and $b$, on each lattice site:
\begin{equation}
H=-\sum_{\expval{ij}} Z_{ia} Z_{ib} Z_{ja} Z_{jb}-g\sum_{i}   X_{ia}  X_{ib}. \label{eq:stabilizer_local}
\end{equation}

This model exhibits a finite temperature phase transition in the 2D Ising universality class, and due to the commuting projector property, the corresponding $T_c$ is exactly same as the Onsager's solution \cite{onsager1944} to the classical Ising model on the square lattice, $H = - \sum_{<i,j>} s_i s_j$, irrespective of the value of $g$. Let us first consider the negativity between one spin on a single site, say, `a' spin on site 0, and the rest of system. As just discussed, to calculate the negativity, we only need the reduced  density matrix for spins at the boundary, which in this case are the spins on sites 0 and four neighbors of site 0. For simplicity, we present the result of the negativity only for a specific range of $g$, namely, $e^{-8\beta} < \tanh(\beta g) < e^{-4\beta }$ where the calculation is technically simpler, see supplemental material for details. This is sufficient to illustrate the singular nature of negativity across the finite temperature transition hinted above. One finds that the negativity $E_N$ is given by:

\begin{eqnarray}
E_N & = & \log\left\{     1-4A \left[    \cosh(\beta g) e^{-4\beta }  - \sinh(\beta g) e^{4\beta}   \right]  \left( 1+4c_1+ \right. \right. \nonumber \\
 & & 	\left. \left. 2c_2+c_3 \right)   \right\}.
\end{eqnarray}
where $A^{-1}=2^5\cosh(\beta g )(\cosh[4](\beta )+\left(  c_1+\frac{1}{2}c_2 \right) \sinh[2](2\beta) +c_3\sinh[4](  \beta  )    )$ and $\{ c_i \vert  i=1,2,3\}$ are given by the expectation values of certain local operators measured by the bulk density matrix $\rho_{\textrm{bulk}}\sim \exp{-\beta(H_A+H_B)}$ so that they are all singular functions of the tuning parameters $g,T$ across the critical point. Inheriting the singularity of $c_i$, the negativity between the single spin and the rest of the system is also singular. Note that there is no symmetry reason for the singularity to cancel out in the particular combination of $c_i$'s that enter the expression for $A$. To confirm this, we calculated the coefficients $c_i$ within the mean field approximation and checked that $E_N$ is indeed singular across the transition.

We can now argue rather generally that negativity will be singular across a phase transition in a commuting projector Hamiltonian for arbitrary bipartition scheme. Due to the property in Eq.\ref{eq:reduced_nega}, one only needs to consider the reduced density matrix $\rho_{\partial A,\partial B}$ for the boundary spins, whose partial transpose takes the form 
\begin{equation}\label{eq:rho_proof}
\rho^{T_B}_{\partial A,\partial B} = \left(e  ^{-\beta H_{AB}}  \right)^{T_{\partial B}}  \sum_m c_m O_m,
\end{equation}
where $O_m$ can be expressed as the tensor product of Pauli matrices acting on the Hilbert space $\partial A, \partial B$. The associated coefficients $ c_m$ are proportional to the expectation value of $O_m$ with respect to the bulk density matrix $\rho_{\textrm{bulk}}$, and are therefore a singular function of the tuning parameter across $T_c$, similar to the coefficients $c_1, c_2, c_3$ discussed above for the case of a single site negativity. From Eq.\ref{eq:rho_proof}, it follows that the negativity is
\begin{equation}
E_N = \log(\sum_m c_m     f_m),
\end{equation}
where $f_m=\sum_{\sigma_A,\sigma_B} \abs{\left(e  ^{-\beta H_{AB}}  \right)^{T_B}   \left(  \sigma_A,\sigma_B  \right)  } O_m(\sigma_A,\sigma_B)$, and $\sigma_A, \sigma_B$ denote the value of the boundary spins in the basis where $H_{AB}$ and all operators $O_m$ are simultaneously diagonalizable (this is always possible since the Hamiltonian is a sum of commuting projectors). In contrast to $c_m$, the coefficients $f_m$ are determined only by the reduced density matrix on the boundary spins via the above expression, and are oblivious to the bulk criticality. Therefore, the negativity inherits the singularity associated with the bulk criticality due to its dependence on coefficients $c_m$.

\begin{figure}
 	\centering
 		\includegraphics[width=0.45\textwidth]{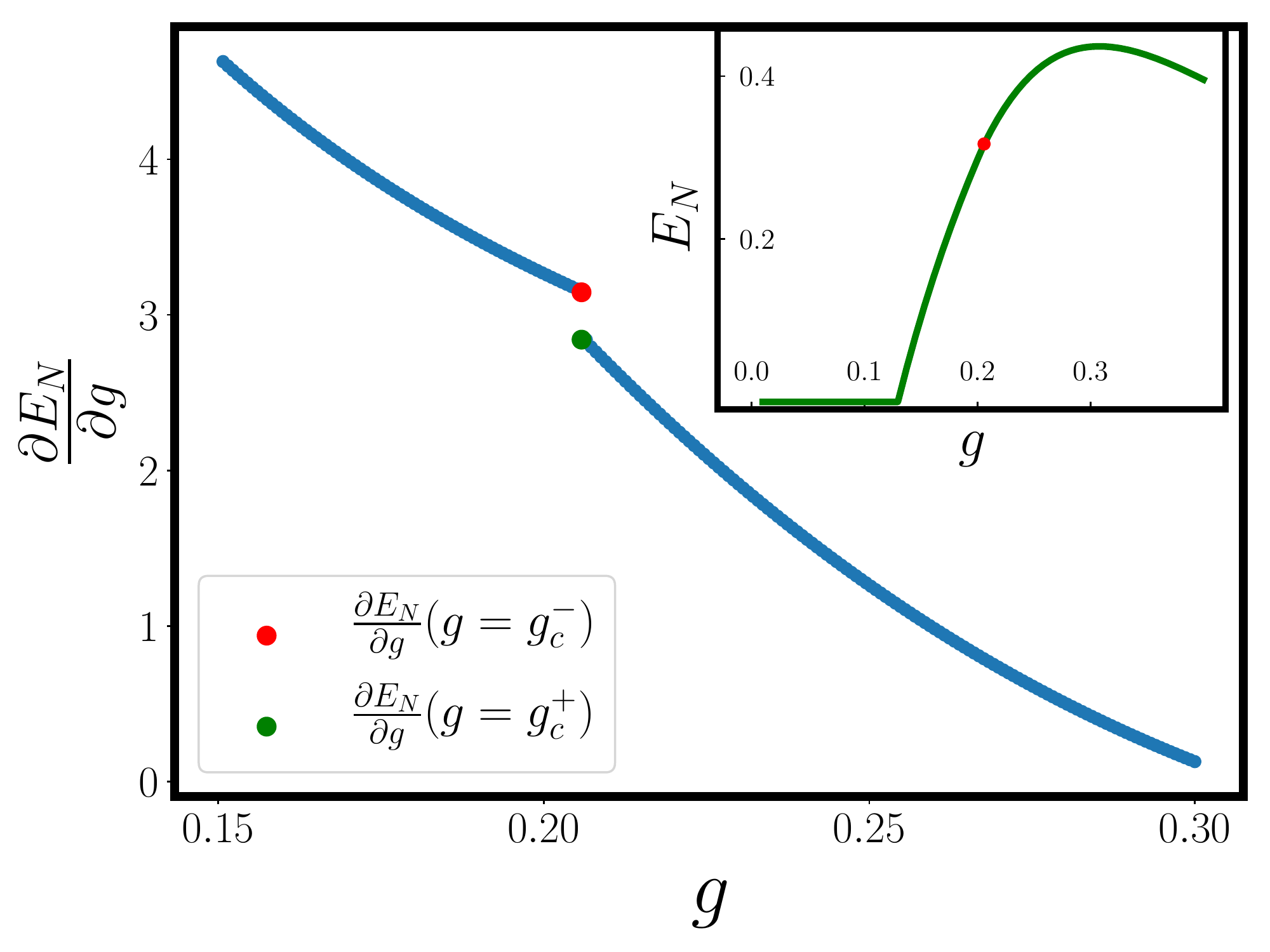}
 		\caption{First derivativie of the negativity as a function of the parameter $g$ (see Eq. \ref{eq:spherical_nonlocal}) at $T=0.15$ for the spherical model. The inset shows the negativity as a function of $g$, where the red dot labels the critical point.}
 		\label{fig:spherical_inf}
 \end{figure}

Finally, we consider a completely different class of models which are also exactly solvable and in which one again finds that the negativity is singular across the phase transition. In particular, consider the quantum spherical model\cite{vojta1996quantum}:
%

\begin{equation}
H=\frac{1}{2} g \sum_{i=1}^N p_i^2 -\frac{1}{2N} \sum_{i,j=1}^N x_ix_j  + \mu \left[ \sum_{i=1}^N x_i^2 -\frac{N}{4} \right], \label{eq:spherical_nonlocal}
\end{equation}
where $x_i$ and $p_j$ satisfy the canonical commutation relation $[x_i,p_j] = i \delta_{ij}$, while the constraint $\left< \sum_{i=1}^N x_i^2 \right> =\frac{N}{4}$ is imposed only on average via the Lagrange multiplier
$\mu$. The above model shows a phase transition associated with spontaneously breaking of the Ising symmetry $x_i \rightarrow - x_i$ at temperature $\beta^{-1}_c$ determined via $\sqrt{g_c} = \frac{1}{2}\tanh\left( \frac{1}{2} \beta_c \sqrt{g_c}  \right)$. In the ordered phase, $\mu$ is pinned to $1/2$. The negativity of this model can be calculated analytically using the correlation matrix technique of Ref.\cite{audenaert2002entanglement}. Dividing the system into two equal halves, one finds that the negativity $E_N = \textrm{Max}\{0, - \log(\nu)\}$ where $\nu=\frac{2}{\beta\sqrt{g}}   \coth(\frac{1}{2}\beta \sqrt{g})$ in the ordered phase, while $\nu = \frac{1}{2} \sqrt{\frac{2\mu-1}{g}} \coth(\frac{1}{2}\beta \sqrt{(2\mu-1)g})$ in the disordered phase where the chemical potential is given by the equation $\sqrt{\frac{2g}{\mu}} =\tanh\left( \frac{1}{2} \beta \sqrt{2g\mu}  \right)$. Using these equations, one finds that the first derivative of the negativity across the phase transition is discontinuous: $
\eval{\frac{\partial E_N}{\partial g}}_{g_c^{+}}= \frac{1}{g_c} +\frac{\beta_c^2}{12} \left(1 - \frac{8}{4 + \beta_c - 4 \beta_c g_c} \right)$ while
 $\eval{\frac{\partial E_N}{\partial g}}_{g_c^{-}}=\frac{4 + \beta_c - 4 \beta_c g_c}{8 g_c}$. As shown in Fig.\ref{fig:spherical_inf}, the first derivative of $E_N$ clearly exhibits a discontinuity at the thermal critical point.

 So far we have showed that finite temperature transitions in quantum systems can show singular features in entanglement negativity, despite the fact that the universal critical exponents associated with these transitions are still given by classical statistical mechanics. Therefore, it is legitimate to ask whether negativity can at all distinguish the spontaneous symmetry breaking at finite temperature with spontaneous symmetry breaking at zero temperature? The answer is in the affirmative. For concreteness, again consider the exactly solvable model in Eq.\ref{eq:inf_stabilizer} although the argument is rather general. Below $T_c$, and in the absence of an infinitesimal symmetry breaking field, the partition function gets equal contribution from both positive and negative values of the order parameter. On the other hand, in the thermodynamic limit, and in the presence of an infinitesimal symmetry breaking field, only one of the two sectors contribute, and therefore, the thermal entropy with and without field satisifies $S(h= 0) - S(h = 0^{+}) = \log(2)$. This is why the spontaneous symmetry breaking at a finite temperature is an example of ergodicity breaking \cite{goldenfeld_book} or relatedly, a `self-correcting classical memory' \cite{bacon2006}. Since this is a classical phenomena, a faithful measure of quantum correlations should be insensitive to it. One may now explicitly calculate the negativity with and without infinitesimal symmetry breaking field for Hamiltonian in Eq.\ref{eq:inf_stabilizer}, and show that $E_N(h = 0) = E_N(h=0^{+})$ (see supplemental material). Schematically, at a mean-field level, $\rho(h=0)= \left(\rho(m*) + \rho(-m*)\right)/2$ where $m*$ is the mean-field value of the order parameter, and therefore $|\rho^{T_B}|_1(h=0) =  \left(|\rho(m*)^{T_B}|_1 + |\rho(-m*)^{T_B}|_1\right)/2 = |\rho(m*)^{T_B}|_1 = |\rho^{T_B}|_1(h=0^{+}) $. In strong contrast, for spontaneous symmetry breaking at $T = 0$, when $ h = 0$, the ground state wavefunction (and not the density matrix) is a sum of the ground state wavefunctions corresponding to positive and negative order parameters (a `cat state') while at $h = 0^{+}$, only one of the two sectors contribute. Therefore, all measures of quantum entanglement, including von Neumann entanglement entropy and in particular negativity satisfy $E_N(h= 0) - E_N(h= 0^{+})= \log(2)$.
 
 The models introduced in this paper allowed for a rather straightforward evaluation of negativity while illustrating non-trivial features. It is natural to wonder whether one can calculate any other measures of mixed state entanglement for similar models. To that end, we now present a result on the entanglement of formation $E_F$, a quantity which is generally rather hard to calculate since it requires optimization over all possible states. Consider the following Hamiltonian which is closely related to the Hamiltonians in Eqs.\ref{eq:H_nonlocal_stab} and \ref{eq:stabilizer_local}:
 
 \begin{equation}
 H=-\frac{1}{2L}  \left( \sum_{i=1}^L  Z_{i1}   Z_{i2}    \right)^2 -g \sum_{i=1}^L  X_{i1}   X_{i2} . \label{eq:H_eof}
 \end{equation}
 This Hamiltonian exhibits a finite temperature phase transition at $T_c = 1$. For defining the entanglement of formation $E_F$, similar to our earlier discussion,  we choose a bipartition that cuts through the two spins 1 and 2 on a chosen site $s$. A straightforward analysis shows (see supplemental material) that in the thermodynamic limit, $E_F$ is exactly given by that corresponding to the mean-field density matrix defined as $\rho_{MF} \propto e^{- \beta H}$ where $H = -m Z_{s1} Z_{s2} - g X_{s1} X_{s2}$ and $m$ satisfies the mean-field equation $\tanh(\beta m) = m$. Using the exact result by Wooters on $E_F$ for two qubits (Ref.\cite{wooters1997}), this yields an analytical expression for $E_F$. Unfortunately, in this model, the entanglement of formation exhibits a sudden death temperature which is lower than $T_c$ for all values of $g$, and therefore, $E_F$ is zero in the vicinity of the transition.
 
 To summarize, we analytically demonstrated that negativity is singular across finite temperature phase transitions for several models. This may seem counterintuitive since the universal properties associated with transitions are controlled by a purely classical Hamiltonian with the same symmetries.  One way to resolve this apparent tension is to note that negativity is sensitive to short-distance quantum correlations close to the bipartition boundary. Since even local properties, such as magnetization or energy density, are singular across the transition, one expects that the area-law associated with negativity will generically pick up a singular contribution as well. In contrast to our results, Ref.\cite{sherman2016}, based on small scale numerics ($L \lesssim 10$ sites), concluded that negativity for the 2+1-D quantum Ising model has no singularity across the finite temperature transition. Although we don't have any analytical results for the negativity of 2+1-D quantum Ising model, for the general reasons just mentioned, we  suspect that negativity will be singular in this model as well. As is evident from the insets of Figs.\ref{fig:meanfield} and \ref{fig:spherical_inf}, it can be rather hard to detect the singularity in negativity unless one has access to an analytical expression, or precise numerical data on very large system sizes. We hope that our results will prompt further in-depth numerical and field-theoretic calculations of entanglement negativity in systems that exhibit finite temperature transitions.
 
The singularity in negativity for the local models discussed in this paper is somewhat analogous to the singular area-law contribution at a zero-temperature QPT discussed in Ref.\cite{metlitski2009}. At the same time, the absence of finite temperature topological order \cite{hastings2011} in our models  suggests that unlike the zero temperature case, there is no additional subleading O(1) constant. If so, then one might be able to cancel out the singular contribution completely via an appropriate subtraction scheme, perhaps similar to that in Ref.\cite{Kitaev06_1}. Relatedly, it would be also interesting to find models where the singularity associated with negativity is universal and unrelated to the classical correlations. On a more practical front, it would be interesting to devise models where the singularity in negativity can be measured experimentally, using quantum state tomography \cite{lanyon2017}, or via swap-based methods on multiple copies of a system \cite{gray2017, islam2015, Kaufman794}. 

\begin{acknowledgments}
 	\emph{Acknowledgments:} 
 We thank John McGreevy and especially Tim Hsieh for helpful discussions and comments on the draft. TG is supported by an Alfred P. Sloan Research Fellowship. This work used the Extreme Science and Engineering Discovery Environment (XSEDE) (see Ref.\cite{xsede}), which is
 supported by National Science Foundation grant number ACI1548562.
 \end{acknowledgments}
  
 \bibliography{v1bib}
 \renewcommand\refname{Reference}
\newpage 

\onecolumngrid

\begin{center}
	\textbf{
		\Large{Supplemental Material}}
	\vspace{0.4cm} 
	
\end{center}

\section{1. General Results Regarding Commuting Projector Hamiltonians}

\subsection{1a. Partial Transposition Preserves the Set of Eigenvectors}\label{appendix_partial_transpose}
Consider a commuting projector Hamiltonian $H=H_A+H_B+H_{AB}$ , where $H_A$ and $H_B$ denote the part of $H$ with support only in real space region $A$ and $B$, and $H_{AB}$ denotes the interaction between $A$ and $B$.  Define $\{O_m  \}$ as the set of local commuting operators, a commuting projector Hamiltonian can be written as $H=\sum_m c_mO_m$. The thermal density matrix, $\rho=e^{-\beta H}/Z$ with $Z=\Tr e^{-\beta H}$, can be expanded as: $\rho= \sum_{\alpha} d_{\alpha}Q_{\alpha}$, where each $\{Q_{\alpha}\}$ is a tensor product of operators from the set $\{O_m\}$.   Since all operators in $H$ commute, $H$, $\rho$, and $\{ O_m\}$ share the same eigenvectors. Under the partial transpose over the Hilbert space in $B$, one obtains $\rho ^{T_B}=\sum_{\alpha}  d_{\alpha} Q_{\alpha}^{T_B}$. If $Q_{\alpha}$
only acts on A or B, then $Q_{\alpha}^{T_B}=Q_{\alpha}$. Only when the support of $Q_{\alpha}$ involves  $A$ and $B$ simultaneously is it possible for $Q_{\alpha }$ to receive a minus sign under partial transpose. This implies that the operators basis for $\rho^{T_B}$ is still $\{Q_{\alpha}  \}$, and thus the eigenvectors of $\rho^{T_B}$ are exactly the same as those of $\rho$, and the eigenvalues of $\rho^{T_B}$ can be obtained by replacing $\{O_{m} \}$ by their eigenvalues. In the argument above we implicitly assumed that all matrix elements of $\{Q_{\alpha}\}$ are real in the basis where we perform a partial transpose. If there exists complex matrix elements instead, $\{Q_{\alpha}\}$ might get a minus sign even when $\{Q_{\alpha}\}$ acts only on $A$ or $B$. Nevertheless, one can check that $\rho^{T_B}$ is still generated by tensor products of $\{O_m\}$, and therefore the conclusion remains the same.

\subsection{1b. Partial Trace Preserves the Set of Eigenvectors}\label{appendix:partial_trace}
Here we show that for commuting projector Hamiltonians, the thermal density matrix $\rho$ and the reduced density matrix $\rho_A$ obtained by tracing out all the degrees of freedom in $B$ share the same set of eigenvectors. As discussed above, $\rho= \sum_{\alpha} d_{\alpha}Q_{\alpha}$, where $\{Q_{\alpha}\}$ collects all possible operators from the product of commuting operators $\{O_m\}$. By tracing out all the degrees of freedom in $B$ for $\rho$, basis operators  in $\{ Q_{\alpha } \}$ which act non-trivially on $B$ vanish. This implies that the operator basis of reduced density matrix $\rho_A$ is generated by the those operators in $\{Q_{\alpha} \}$ which act on $B$ trivially, and thus $\rho_A$ commutes with all local commuting operators.

\subsection{1c. Bipartite Negativity from Reduced Density Matrix on Boundary}\label{append:boundary_nega}
Here we show that the negativity between two spatial regions of a thermal density matrix of a commuting projector Hamiltonian equals the negativity of the reduced density matrix localized on  the boundary of the bipartition. Following the notation in the main text, we define $\partial A(\partial B)$ as collection of spins on the boundary of $A(B)$ that interacts with $B(A)$, and define $A'(B')$ as the collection of spins in the bulk of $A(B)$ that only couples to spins in $A(B)$. We decompose a commuting projector Hamiltonian as $H = H_A + H_B + H_{AB}$, so that $H_A(H_B)$ denotes the interaction between the bulk spins in $A(B)$, and $H_{AB}$ denotes the interaction between the boundary spins in $\partial  AB=\partial A\bigcup \partial B$. For simplicity, we also assume that the system is time reversal invariant, so that  for $\rho =e^{-\beta H}/Z$, the partial transpose over  the Hilbert space in $B$ acts non-trivially only on $H_{AB}$:
\begin{equation}
\left(\rho \right)^{T_B}=\frac{1}{Z} \left( e^{-\beta H_{AB}}   \right)^{T_{\partial B}} e^{-\beta\left( H_{A}+H_B  \right)  } .
\end{equation} 
As discussed above, a partial transposed density matrix is still generated by local commuting operators which are present in $H$. This implies that one can find a common eigenbasis for $H_A, H_B$ and $H_{AB}^{T_{\partial B}}$, and the eigenvalues of $\rho^{T_B}$ can be obtained by replacing all local operators by their eigenvalues. Consequently, the eigenvalues of  $\rho ^{T_B}$ take the form:
\begin{equation}
\lambda=\frac{1}{Z} \left( e^{-\beta H_{AB}}   \right)^{T_{\partial B}}\left(\sigma_A,\sigma_B\right) e^{-\beta\left( H_{A}(s_A,\sigma_A)+H_B(s_B,\sigma_B)  \right)  }.
\end{equation} 
where $s_{A(B)}$ denotes the spin configuration in the bulk of $A(B)$, and $\sigma_{A(B)}$ denotes the spin configuration on the boundary that interact with $B(A)$. The one-norm of $\rho^{T_B}$ can be obtained by summing all the absolute values of eigenvalues:

\begin{equation}\label{eq:prove_one_norm}
\norm{\rho^{T_B} }_1 =\frac{      \sum_{\sigma_A,\sigma_B}  \abs{ \left( e^{-\beta H_{AB}}   \right)^{T_B}\left(\sigma_A,\sigma_B\right)     }    \sum_{s_A,s_B}  e^{-\beta\left( H_{A}(s_A,\sigma_A)+H_B(s_B,\sigma_B)  \right)  }         }{         \sum_{\sigma_A,\sigma_B} \left( e^{-\beta H_{AB}\left(\sigma_A,\sigma_B\right)  }   \right)          \sum_{s_A,s_B}  e^{-\beta\left( H_{A}(s_A,\sigma_A)+H_B(s_B,\sigma_B)  \right)  }     }.
\end{equation}
The observation that partial transpose only affects the operators on the boundary motivates us to consider the reduced density matrix on the boundary:
\begin{equation}
\rho_{\partial AB} = \frac{1}{Z} \Tr_{A',B'}  e^{-\beta H} = \frac{1}{Z} e^{-\beta H_{AB}}   \Tr_{A',B'}     e^{-\beta \left(H_{A}+H_B \right)  }  .
\end{equation}
We take the partial transpose over $\partial B$
\begin{equation}
\left( \rho_{\partial AB} \right)^{T_{\partial B}} 
=  \frac{1}{Z} \left\{   e^{-\beta H_{AB}} \right\}^{T_{\partial B}}   \Tr_{B'}     e^{-\beta H_B   }  \Tr_{A'}     e^{-\beta H_A   },
\end{equation}
where the commutative property of each local operator is used. As a result, the eigenvalue of $\left( \rho_{\partial AB} \right)^{T_{\partial B}}$ is:
\begin{equation}\label{eq:reduced_eig}
\lambda_{\sigma_A,\sigma_B}   = \frac{1}{Z} \left( e^{-\beta H_{AB}}   \right)^{T_{\partial B}}\left(\sigma_A,\sigma_B\right)         \sum_{s_A,s_B}  e^{-\beta\left( H_{A}(s_A,\sigma_A)+H_B(s_B,\sigma_B)  \right)  } .
\end{equation}
By summing all absolute values of $\lambda_{\sigma_A, \sigma_B}$ for $\norm{\rho_{\partial AB}^{T_{\partial B}} }_1 $ and comparing it with Eq.\ref{eq:prove_one_norm}, one finds that 

\begin{equation}
\norm{\rho^{T_B} }_1 =\norm{\rho_{\partial AB}^{T_{\partial B} }}_1,
\end{equation}
which implies that the negativity of two spatial regions is given by the boundary of those two spatial regions. In fact with a similar calculation, one can show that the above equality also holds true for any commuting project Hamiltonian without time reversal symmetry.

\section{2. Calculational details of negativity for various models discussed in the main text}

\subsection{2a. Infinite-Range Commuting Projector Hamiltonian}
Consider a one-dimensional lattice of size $L$ where each lattice site has four qubits, the model Hamiltonian is 
\begin{equation}
\begin{split}
H=&-\frac{1}{4L}  \left( \sum_{i=1}^L  \left(Z_{i1}   Z_{i2} + Z_{i3} Z_{i4}  \right) \right)^2 -g_z \sum_{i=1}^L  Z_{i1}   Z_{i2} Z_{i3} Z_{i4}  -g_x \sum_{i=1}^L \left( X_{i1}   X_{i2}+X_{i3} X_{i4}  \right).
\end{split}
\end{equation}	
The density matrix at inverse temperature $\beta$ is $\rho=\frac{1}{Z}e^{-\beta H}$ with $Z=\Tr e^{-\beta H}$. Since every local term commutes, we can perform Hubbard-Stratonovich transformation for $e^{-\beta H}$:

\begin{equation}\label{eq:ir_thermal_state}
e^{-\beta H} =\sqrt{\frac{  \beta  L}{ \pi   }} \int dm e^{ -\beta L m^2     -     \beta \sum_{i=1 }^L H_i(m)},
\end{equation}
where a local Hamiltonian $H_i(m)$ for $i$-site of four spins is defined as :
\begin{equation}
H_i(m)= - m  (  Z_{i1}   Z_{i2} + Z_{i3} Z_{i4}   ) -   g_z  Z_{i1}   Z_{i2} Z_{i3} Z_{i4}  -g_x  \left(X_{i1}   X_{i2}+X_{i3} X_{i4}  \right).
\end{equation}
Eq.\ref{eq:ir_thermal_state} implies that all sites are separable since $\rho$  manifestly takes the form $\rho=\sum_k p_k \rho^1_k \otimes \cdots\otimes \rho_k^L$ where $p_k\geq 0 $, $\rho_k^i$ is a local density matrix on $i$-th site. As a result, to have non-zero negativity, an entanglement cut should be made across one of the sites (say $s$-th site) such that four spins on $s$-th site are not in the same subsystem. In the following calculation, $A$ comprises all the lattice sites with site index $i<s$ and two spins labelled by $1,3$ on $s$-th site while $B$ comprises all the lattice sites with site index $i>s$ and two spins labelled by $2,4$ on $s$-th site. 
The negativity $E_N$ can be calculated via a replica trick:
\begin{equation}\label{eq:inf_nega}
E_N=\log \norm{\rho^{T_B}}_1 = \lim_{n_{e}\to 1} \frac{\Tr\left[ \left( ( e^{-\beta H}  )^{T_B} \right)^{n_e}    \right]}{\Tr\left[e^{-\beta H}    \right]}.
\end{equation}
Notice that $n_e$ is an even number as performing trace, but analytic continuation $n_e\to 1$ is taken in the end.
First we  calculate the thermal partition function:
\begin{equation}\label{eq:inf_stabilizer_Z}
Z=\Tr{ e^{-\beta H} } =\left(\frac{\beta L}{\pi}  \right)^{  \frac{1}{2} }   \int dm e^{-\beta L m^ 2} \Tr{e^{-\beta   \sum_{i=1}^L H_i(m)   }    }  = \left(\frac{\beta L}{\pi}  \right)^{  \frac{1}{2} }   \int dm e^{-\beta L f(m)	    }
\end{equation}
where 
\begin{equation}
\beta f(m)= m^2 - \log \left[e^{\beta g_z} \cosh(2 \beta m )  +e^{-\beta g_z}    \right]   -\log \left[  8 \cosh[2](\beta g_x)   \right].
\end{equation}
The integral over $m$ is dominated by the saddle point $m^*$, which satisfies $\eval{\frac{\partial f(m)}{\partial m}}_{m^*} =0$: 
\begin{equation}\label{eq:inf_stabilizer_saddle}
\frac{\sinh(2\beta m^*)}{  \cosh(2\beta m^* )   +e^{-2\beta g_z}  } =m^*.
\end{equation}
The critical behavior of $m^*$ can be determined by expanding Eq.\ref{eq:inf_stabilizer_saddle} to $O(m^{*3})$:
\begin{equation}
\frac{2\beta m^*}{1+w} +\frac{4(w-2)}{3\left( 1+w   \right)^2}  \beta^2m^{*3}=m^*,
\end{equation}
where $w(\beta)\equiv e^{-2\beta g_z}$. 
Define $\beta_c\equiv \frac{1+w(\beta_c)}{2}$, for $\beta>\beta_c$, we can have non-zero solution for $m^*=\pm m_0$:

\begin{equation}
m_0=\sqrt{ \frac{3\beta_c\left(  \beta-\beta_c \right)}{\beta^3\left( 3-2\beta_c  \right)} } 
\sim \sqrt{T_c-T}
\end{equation}
while for $\beta<\beta_c$, $m^*=0$ is the only allowed solution. 
Notice that the critical inverse temperature $\beta_c $ is determined by solving the transcendental equation:
\begin{equation}
2\beta_c=1+e^{-2\beta_cg_z}.
\end{equation}
On the other hand, for the calculation of $\Tr\left[ \left( ( e^{-\beta H}  )^{T_B} \right)^{n_e}    \right]$, since each site are separable, taking partial transpose over $B$ amounts to only taking the partial transpose on the two spins labelled by $2,4$ on the  $s$-th site:

\begin{equation}\label{eq:if_transposed_trace}
\left[  e^{-\beta H}  \right]^{T_B} =\sqrt{\frac{  \beta  L}{ \pi   }} \int dm e^{ -\beta L m^2     -     \beta \sum_{i \neq  s } H_i(m)}   \left[ e^{  -\beta  H_s(m)  }\right]^{T_B}  .
\end{equation}
By introducing $n_e$ replicas, we have
\begin{equation}\label{eq:nth_moment}
\begin{split}
\Tr{  \left[ \left(  e^{-\beta H} \right) ^{T_B}   \right]^{n_e}} &=\left(\frac{\beta L}{\pi}\right)^{\frac{n_e}{2}}\int \prod_{a=1}^{n_e} dm_{a} e^{-\beta L\sum_{a=1}^{n_e}  m_a^ 2} \Tr_{i\neq s}\left\{  e^{-\beta\sum_{a=1}^{n_e} \sum_{i\neq s} H_i(m_a)   }    \right\} \Tr_{s}\left\{ \prod_{a=1}^{n_e} \left[  e^{-\beta H_s (m_a)} \right]^{T_B}   \right\}\\
&=\left(\frac{\beta L}{\pi}\right)^{\frac{n_e}{2}}\int \prod_{a=1}^{n_e} dm_{a}   e^{-\beta LF_{n_e}(\{ m_a\}  )}  \frac{ \Tr_{s} \left\{\prod_{a=1}^{n_e} \left[  e^{-\beta H_s (m_a)} \right]^{T_B} \right\}}{ \Tr_{s}\left\{ \prod_{a=1}^{n_e}  e^{-\beta H_s (m_a)} \right\}}
\end{split}
\end{equation}
where 
\begin{equation}
\beta F_{n_e}(\{m_a  \}  )= \sum_{a=1}^n m_a^2 - \log \left[e^{\beta n_eg_z} \cosh(2 \beta  \sum_{a=1}^{n_e}  m_a )  +  e^{-\beta n_e g_z}    \right]   -\log \left[  8 \cosh[2](\beta  n_eg_x)   \right].
\end{equation}
This multi-dimensional integral is again dominated by saddle points  $\{m_a^*\vert a=1,2,\cdots, n_e  \}$, which can be obtained from $\eval{\frac{\partial F_{n_e}(\{m_a  \} )}{\partial m_a}}_{m_a^*} =0$: 
\begin{equation}
\frac{\sinh(2\beta \sum_{a=1}^{n_e}m^*_a)}{  \cosh(2\beta\sum_{a=1}^{n_e} m^*_a )   +e^{-2\beta n_eg_z}  } =m^*_a   \quad \forall a.
\end{equation}
Assuming replica symmetry is preserved, we have $m^*_{n_e}=m^*_a~ \forall a$ with
\begin{equation}
\frac{\sinh(2n_e\beta m^*_{n_e})}{  \cosh(2n_e\beta m^*_{n_e} )   +e^{-2\beta n_eg_z}  } =m^*_{n_e}.
\end{equation}
As $n_e\to 1$, the above equation is exactly the saddle point equation for the thermal partition function (Eq.\ref{eq:inf_stabilizer_saddle}). This implies $\lim_{n_e\to 1} m^*_{n_e} = m^*$.
By plugging Eq.\ref{eq:inf_stabilizer_Z} and Eq.\ref{eq:nth_moment} into Eq.\ref{eq:inf_nega}, one finds
\begin{equation}
\norm{\rho^{T_B}}_1=  \frac{       \int dm e^{- \beta Lf(m,g_z,g_x)}    \norm{\rho^{T_B}_s(m)}_1     }{   \int dm e^{- \beta Lf(m,g_z,g_x)} },
\end{equation}
where 
\begin{equation}
\rho_s(m)\equiv \frac{  e^{-\beta H_s(m)}}{   \Tr_s\left\{    e^{-\beta H_s(m)}\right\}}. 
\end{equation}
For $T>T_c$, there is an unique saddle point $m^*$, and 
\begin{equation}
\norm{\rho^{T_B}}_1= \norm{\rho^{T_B}_s(m^*)}_1    \frac{    \int dm e^{- \beta Lf(m,g_z,g_x)}        }{   \int dm e^{- \beta Lf(m,g_z,g_x)} }=   \norm{\rho^{T_B}_s(m^*)}_1. 
\end{equation}
For $T<T_c$, there are two saddle points $m^*=\pm m_0$, and thus we arrive at 
\begin{equation}
\norm{\rho^{T_B}}_1=       \frac{ \norm{\rho^{T_B}_s(m_0)}_1     \int_{\text{around}~ m_0} dm e^{- \beta Lf(m,g_z,g_x)}       +\norm{\rho^{T_B}_s(-m_0)}_1      \int_{\text{around}~ -m_0} dm e^{- \beta Lf(m,g_z,g_x)}       }{        \int_{\text{around} ~m_0} dm e^{- \beta Lf(m,g_z,g_x)}    +  \int_{\text{around} ~-m_0} dm e^{- \beta Lf(m,g_z,g_x)}     }.
\end{equation}
Since $\norm{\rho^{T_B}_s(m_0)}_1 $ = $\norm{\rho^{T_B}_s(-m_0)}_1 $, we have 
\begin{equation}
\norm{\rho^{T_B}}_1= \norm{\rho^{T_B}_s(m^*)}_1. 
\end{equation}
This result implies that to calculate the bi-partite negativity between $A$ and  $B$, it is sufficient to calculate the reduced density matrix for $s$-th site ($\rho_s$) where we made an entanglement cut.  Incidentally, the above calculation explicitly demonstrates the claim  $E_N(h = 0) = E_N(h=0^{+})$ mentioned in the main text where  $E_N(h = 0)$ is the negativity in the absence of an infinitesimal symmetry breaking field (so that it receives contribution from both $m_0$ and $-m_0$) while  $E_N(h=0^{+})$ is the negativity in the presence of such a field so that it receives contribution only from one saddle point (say, $m_0$). 
From now on, we suppress lattice site index $s$ in the calculation since only four qubits on a single site is relevant. Meanwhile, $m$ will replace $m^*$ as the mean-field order parameter for brevity. The local density matrix is

\begin{equation}
\rho_{s}=\frac{1}{Z_{s}}e^{-\beta H_{s}}=\frac{1}{Z_{s}}  e^{    \beta m (  Z_{1}   Z_{2} + Z_{3} Z_{4}   ) +  \beta   g_z   Z_{1}   Z_{2} Z_{3} Z_{4} +\beta  g_x  \left(X_{1}   X_{2} +X_{3} X_{4} \right)      },
\end{equation}
where the partition function $Z_s$ is 
\begin{equation}
Z_s=  \Tr e^{-\beta H_s}=   8\left(   \cosh(\beta g_x)\right)^2 \left( e^{\beta g_z}\cosh(2\beta m ) +e^{-\beta g_z }  \right)
\end{equation}
By taking partial transpose over $\{ 2,4\} \in B$, we have 
\begin{equation}
\begin{split}
\left( e^{-\beta H_{s}}  \right)^{T_{24}}& =   e^{\beta g_z Z_1Z_2Z_3Z_4} \left[ (\cosh(\beta g_x))^2  e^{\beta m(Z_1Z_2+Z_3Z_4)}  +   (\sinh(\beta g_x))^2  e^{- \beta m(Z_1Z_2+Z_3Z_4)}    X_1X_2X_3X_4      \right] \\
&  +\frac{1}{2} \sinh(2\beta g_x)     e^{-\beta g_z Z_1Z_2Z_3Z_4} \left[   e^{\beta m (-Z_1Z_2+Z_3Z_4)  }X_1X_2 +  e^{\beta m (Z_1Z_2-Z_3Z_4)  }X_3X_4      \right].
\end{split}
\end{equation}
Due to the simple form of $\left( e^{-\beta H_{s}}  \right)^{T_{24}}$,  we are able to obtain all the eigenvalues of $\rho_s^{T_{24}}$, and exploit the following formula to calculate the negativity:

\begin{equation}\label{eq:nega}
E_N=\log \left[\sum_{i} \abs{\nu_i}   \right] = \log \left[1-2\sum_{\nu_i<0} \nu_i  \right],
\end{equation}
where $\{\nu_i \} $ denotes eigenvalues of $\rho_s^{T_{24}}$.
Since $Z_1Z_2,~Z_3Z_4,~X_1X_2,~X_3X_4 $ commute with each other, the corresponding eigenvalues of these operators $z_{12},~z_{34},~x_{12},~x_{34}=\pm 1 $ completely specify an eigenvector of $\left( e^{-\beta H_{s}}  \right)^{T_{24}}$, which takes the following form 

\begin{equation} 
\ket{\psi }=\frac{1}{2}   \left( \ket{s_1,s_2  } \pm \ket{-s_1,-s_2  }   \right) \otimes     \left( \ket{s_3,s_4  } \pm \ket{-s_3,-s_4  }   \right).
\end{equation}
with $s_i=\pm1 $ for $i=1,2,3,4$.  With this observation, the eigenvalues of $\left( e^{-\beta H_{s}}  \right)^{T_{24}}$ can be obtained by replacing operators by their eigenvalues:

\begin{equation}
\begin{split}
\lambda(z_{12},z_{34},x_{12},x_{34})=&e^{\beta g_z z_{12}z_{34} } \left[ (\cosh(\beta g_x))^2  e^{\beta m(z_{12}+z_{34}  )}  +   (\sinh(\beta g_x))^2  e^{- \beta m(z_{12}+z_{34})}   x_{12}x_{34}      \right] \\
&  +\frac{1}{2} \sinh(2\beta g_x)     e^{-\beta g_z z_{12} z_{34}} \left[   e^{\beta m (-z_{12} +z_{34})  }x_{12} +  e^{\beta m (z_{12}- z_{34})  }x_{34}       \right].
\end{split}
\end{equation}
For $T>T_c$, $ m=0$, one finds 
\begin{equation}
\begin{split}
\lambda(z_{12},z_{34},x_{12},x_{34})=&  e^{\beta g_z z_{12} z_{34}} \left[ (\cosh(\beta g_x))^2  +   (\sinh(\beta g_x))^2   z_{12} z_{34}      \right]  +\frac{1}{2} \sinh(2\beta g_x)     e^{-\beta g_z z_{12} z_{34}} \left[  x_{12} +  x_{34}      \right].
\end{split}
\end{equation}
When
\begin{equation}\label{eq:combination}
\begin{cases}
z_{12}=1,~z_{34}=-1,~x_{12}=-1,~x_{34} =-1\\
z_{12}=-1,~z_{34}=1,~x_{12}=-1,~x_{34} =-1,
\end{cases}
\end{equation}
we can have negative $\lambda$: 
\begin{equation}
\lambda=e^{-\beta g_z}\cosh\left(2\beta g_x  \right) -e^{\beta g_z} \sinh(2\beta g_x) .
\end{equation}
Thus, for $T>T_c$, the two-fold degenerate negative eigenvalue of $\rho^{T_{24}}_s$ is 
\begin{equation}
\nu=\frac{e^{-\beta g_z}\cosh\left(2\beta g_x  \right) -e^{\beta g_z} \sinh(2\beta g_x) }{  16\left(   \cosh(\beta g_x)\right)^2  \cosh\left(\beta g_z   \right)},
\end{equation}  
and the negativity can be obtained by using Eq.\ref{eq:nega} :
\begin{equation}
E_N  =\log \left[   1+\max\left\{0,\frac{e^{\beta g_z} \sinh(2\beta g_x)-e^{-\beta g_z}\cosh\left(2\beta g_x  \right)  }{  4\left(   \cosh(\beta g_x)\right)^2  \cosh\left(\beta g_z   \right)}    \right\}      \right].
\end{equation}
Note that at $T_c$, one requires
\begin{equation}
e^{-2\beta_c g_z} < \tanh(2\beta_cg_x) 
\end{equation}
to have non-zero negativity. This is always achievable by tuning $g_x$ since $\beta_c$ is only determined by $g_z$.
For $T<T_c$, depending on the values of $m$, there could be more choices of $(z_{12},z_{34},x_{12},x_{34})$ that can give negative eigenvalues of $\rho_s^{T_{24}}$. For simplicity, we consider $T\to T_c^{-}$, where $ m \sim \sqrt{T_c-T} \to 0^+$, and only the configurations in Eq.\ref{eq:combination} can possibly give negative eigenvalues. This is sufficient for our purpose since we only concern the possibly non-analytic behavior of the negativity. Therefore, as  $T\to T_c^{-}$, the two-fold degenerate negative eigenvalue of $\rho_s^{T_{24}}$ is
\begin{equation}
\nu=\frac{e^{-\beta g_z}\cosh\left(2\beta g_x  \right) -e^{\beta g_z} \sinh(2\beta g_x) \cosh(2\beta m) }{  8\left(   \cosh(\beta g_x)\right)^2 \left( e^{\beta g_z}\cosh(2\beta m ) +e^{-\beta g_z }  \right)}.
\end{equation} 
Finally, the negativity valid for $T>T_c^-$ is given by
\begin{equation}
\boxed{
	E_N=\log \left[   1+\max\left\{0,\frac{e^{\beta g_z} \sinh(2\beta g_x)\cosh(2\beta m)-e^{-\beta g_z}\cosh\left(2\beta g_x  \right)  }{  2\left(   \cosh(\beta g_x)\right)^2  \left(e^{\beta g_z}\cosh(2\beta m ) +e^{-\beta g_z}    \right) }   \right\}      \right] } 
\end{equation}
Due to the singular behavior of $m(T)$:
\begin{equation}
m=\begin{cases}
a\sqrt{T_c-T}   \quad \text{for }\quad T\to T_c^-\\
0  \quad \text{for }  \quad T>T_c,
\end{cases}
\end{equation}
the negativity $E_N$ is also a singular function across $T_c$.

\subsection{2b. Two dimensional Commuting Projector Hamiltonian}
Consider a two dimensional lattice, where each sites has two spins labelled by `a' and `b' respectively, the model Hamiltonian is 
\begin{equation}
H=-\sum_{\expval{ij}} \widetilde{z}_i  \widetilde{z}_j\  -g\sum_{i} \widetilde{x}_i,
\end{equation}
where $\widetilde{z}_i\equiv Z_{ia}Z_{ib}, \widetilde{x}_i\equiv X_{ia}X_{ib}$.  Consider a thermal density matrix $\rho_T\sim \exp{-\beta H}$, here we present the calcualtion of the negativity between one spin on a single site, say, `a' spin in site 0  (subsystem $A$), and its complement (subsystem $B$). As discussed above, to calculate the negativity, we only need the reduced density matrix for spins at the boundary  which in this case are the spins at site 0 and its neighboring sites (labelled as 1,2,3,4 clockwise). The corresponding reduced density matrix on these five sites is
\begin{equation}
\begin{split}
\rho=&A' e^{-\beta g \left(   \widetilde{x}_1 +\widetilde{x}_2+\widetilde{x}_3+\widetilde{x}_4 \right)  } \bigg[  \cosh(\beta g )    e^{\beta \widetilde{z}_0 \left(   \widetilde{z}_1+\widetilde{z}_2+\widetilde{z}_3+\widetilde{z}_4 \right)} + \sinh(\beta g)  e^{\beta \widetilde{z}_0 \left(  \widetilde{z}_1+\widetilde{z}_2+\widetilde{z}_3+\widetilde{z}_4  \right)}  \widetilde{x}_0    \bigg] \\
& \left[  1+c_1\left(   \widetilde{z}_1 \widetilde{z}_2 + \widetilde{z}_2 \widetilde{z}_3 + \widetilde{z}_3 \widetilde{z}_4 + \widetilde{z}_4 \widetilde{z}_1  \right)     +    c_2  \left(     \widetilde{z}_1  \widetilde{z}_3 +  \widetilde{z}_2  \widetilde{z}_4        \right) + c_3   \widetilde{z}_1  \widetilde{z}_2  \widetilde{z}_3  \widetilde{z}_4  \right] .
\end{split}
\end{equation}
Here  $A'$ is determined by demanding $\Tr\rho=1$ and $c_1=\expval{\widetilde{z}_j\widetilde{z}_{j+1}}$;  $c_2=\expval{\widetilde{z}_j\widetilde{z}_{j+2}}$; $c_3=\expval{\widetilde{z}_1\widetilde{z}_2 \widetilde{z}_3\widetilde{z}_4 } $, where the expectation values are taken with respect to the bulk thermal density matrix 
$\rho_{\text{bulk}}\sim \exp{-\beta (H_A+H_B)}$. In fact, due to the property of commuting local terms, $c_i$ can be obtain by considering the thermal state of a bulk classical Hamiltonian,i.e. $g=0$,  with one spin per site, and one just need to replace the composite operator $\widetilde{z}_i$ by a Pauli Z operator at site $i$ (i.e. $Z_i$). For instance,
\begin{equation}\label{eq:append_equiv}
c_1=\expval{\widetilde{z}_j\widetilde{z}_{j+1}}  = \frac{ \Tr{ \widetilde{z}_j \widetilde{z}_{j+1}    e^{\beta \sum_{\expval{ij}} \widetilde{z}_i  \widetilde{z}_j\  +\beta g\sum_{i} \widetilde{x}_i } }    }{\Tr 	e^{\beta \sum_{\expval{ij}} \widetilde{z}_i  \widetilde{z}_j\  +\beta g\sum_{i} \widetilde{x}_i }  }  =    \frac{ \Tr{   Z_jZ_{j+1}    e^{\beta \sum_{\expval{ij}}  Z_i Z_j }  }  }{  \Tr  e^{\beta \sum_{\expval{ij}}  Z_i Z_j }   }  
\end{equation} 
Under the partial transposition over $B$, the density matrix is 

\begin{equation}
\begin{split}
\rho^{T_B}=&A' e^{-\beta g \left(   \widetilde{x}_1 +\widetilde{x}_2+\widetilde{x}_3+\widetilde{x}_4 \right)  } \bigg[  \cosh(\beta g)    e^{\beta \widetilde{z}_0 \left(   \widetilde{z}_1+\widetilde{z}_2+\widetilde{z}_3+\widetilde{z}_4 \right)} + \sinh(\beta g)  e^{   - \beta \widetilde{z}_0 \left(  \widetilde{z}_1+\widetilde{z}_2+\widetilde{z}_3+\widetilde{z}_4  \right)    }  \widetilde{x}_0    \bigg] \\
& \left[  1+c_1\left(   \widetilde{z}_1 \widetilde{z}_2 + \widetilde{z}_2 \widetilde{z}_3 + \widetilde{z}_3 \widetilde{z}_4 + \widetilde{z}_4 \widetilde{z}_1  \right)     +    c_2  \left(     \widetilde{z}_1  \widetilde{z}_3 +  \widetilde{z}_2  \widetilde{z}_4        \right) + c_3   \widetilde{z}_1  \widetilde{z}_2  \widetilde{z}_3  \widetilde{z}_4  \right] ,
\end{split}
\end{equation}
The eigenvalues of $\rho^{T_B}$ can be obtained by just replacing $\widetilde{x}_i, \widetilde{z}_i$ by $\pm 1$. In fact, $e^{-\beta g \left(   \widetilde{x}_1 +\widetilde{x}_2+\widetilde{x}_3+\widetilde{x}_4 \right)  }$ is irrelevant since it just provides a multiplicative factor when summing negative eigenvalues, which got cancelled out by the normalization factor. Effectively, it is sufficient to consider the eigenvalues
\begin{equation}
\begin{split}
\lambda =&A \bigg[  \cosh(\beta g)    e^{\beta \widetilde{z}_0 \left(   \widetilde{z}_1+\widetilde{z}_2+\widetilde{z}_3+\widetilde{z}_4 \right)} + \sinh(\beta g)  e^{   - \beta \widetilde{z}_0 \left(  \widetilde{z}_1+\widetilde{z}_2+\widetilde{z}_3+\widetilde{z}_4  \right)    }  \widetilde{x}_0    \bigg] \\
& \left[  1+c_1\left(   \widetilde{z}_1 \widetilde{z}_2 + \widetilde{z}_2 \widetilde{z}_3 + \widetilde{z}_3 \widetilde{z}_4 + \widetilde{z}_4 \widetilde{z}_1  \right)     +    c_2  \left(     \widetilde{z}_1  \widetilde{z}_3 +  \widetilde{z}_2  \widetilde{z}_4        \right) + c_3   \widetilde{z}_1  \widetilde{z}_2  \widetilde{z}_3  \widetilde{z}_4  \right] ,
\end{split}
\end{equation}
where $\widetilde{x}_0$ and each $\widetilde{z}_i$ takes $\pm 1$, which gives $2^6=64$ eigenvalues, and $A$ is chosen such that the sum of these 64 eigenvalues remains unity.
$\left[ 1+c_1\cdots      \right]$ part is always non-negative since it is obtained by performing partial trace for a density matrix ( positive semidefinite ). As a result, $\lambda$ can be negative only when $ \widetilde{x}_0=-1$ and $e^{2 \beta  \widetilde{z}_0    \left(  \widetilde{z}_1+\widetilde{z}_2+\widetilde{z}_3+\widetilde{z}_4  \right) } < \tanh(\beta g)$.
For a given $g$, there are many choices of $\widetilde{z}_i$ that can result in negative eigenvalues. As our purpose is to check whether the negativity picks up an singularity at a thermal critical point, it is sufficient to restrict $g$ in a range such that only a few eigenvalues are negative. We set $g$ in the range $e^{-8\beta} < \tanh(\beta g) < e^{-4\beta }$, and there are only two negative eigenvalues given by

\begin{equation}
\begin{cases}
&\widetilde{z}_0=1, ~~~ \widetilde{z}_1= \widetilde{z}_2= \widetilde{z}_3= \widetilde{z}_4=-1   \\
&\widetilde{z}_0=-1, ~~~  \widetilde{z}_1= \widetilde{z}_2= \widetilde{z}_3= \widetilde{z}_4=1 .
\end{cases}
\end{equation}
Finally, as 
\begin{equation}
e^{-8\beta} < \tanh(\beta g) < e^{-4\beta }, 
\end{equation}
we obtain the expression of the negativity:

\begin{equation}\label{eq:negativity_single}
\boxed{
	E_N=\log\left\{     1-4A \left[    \cosh(\beta g) e^{-4\beta }  - \sinh(\beta g) e^{4\beta}   \right]  \left( 1+4c_1+2c_2+c_3 \right)   \right\}
}.
\end{equation}

\begin{equation}
A^{-1}=2^5\cosh(\beta g )\left[  \cosh[4](\beta )+\left(  c_1+\frac{1}{2}c_2 \right) \sinh[2](2\beta) +c_3\sinh[4](  \beta  )    \right]
\end{equation}
Due to the singularity of $c_i$ at the thermal critical point, the negativity $E_N$ is expected to be singular. To confirm this intuition, we now adopt a mean-field approach to calculate the coefficient $c_1,c_2,c_3$. The exact nature of singularities associated with $c_i$ for our model would of course be determined by the critical exponents of the 2D Ising model. As shown in Eq.\ref{eq:append_equiv}, $c_i$ is exactly given by the corresponding classical Hamiltonian with one spin per site. As a result, we consider the mean-field Hamiltonian 
\begin{equation}
H=-(3m+Z_0) \left(  Z_1+Z_2+ Z_3+ Z_4 \right),
\end{equation}
we determine $m$ from $m=\expval{Z_i}=\Tr{\rho Z_i}$ for $i=1$ to $4$, where $\rho$ is a density matrix associated with $H$. It is straightforward to obtain the mean-field equation for $m$:
\begin{equation}
m=\frac{ \cosh[4](\beta (3m+1))   \tanh(\beta (3m+1))  +  \cosh[4](\beta (3m-1))   \tanh(\beta (3m-1))        }{\cosh[4](\beta (3m+1))  +  \cosh[4](\beta (3m-1))  }.
\end{equation}
$T_c$ can be determined from this equation, and it is straightforward to show that $m=0$ as $T\to T_c^+$, and $m \sim  \sqrt{T_c-T}$ as $T\to T_c^-$.
Finally, $c_1,c_2,c_3$ can be obtained:
\begin{equation}
\begin{split}
&c_1=c_2=\expval{Z_1Z_2} =\frac{ \cosh[2](\beta (3m+1))   \sinh[2](\beta (3m+1))  +  \cosh[2](\beta (3m-1))   \sinh[2](\beta (3m-1))        }{\cosh[4](\beta (3m+1))  +  \cosh[4](\beta (3m-1))  }\\
&c_3=\expval{Z_1Z_2Z_3Z_4} = \frac{  \sinh[4](\beta (3m+1))  +  \sinh[4](\beta (3m-1))        }{\cosh[4](\beta (3m+1))  + \cosh[4](\beta (3m-1))  }
\end{split}
\end{equation}
Plug the coefficients into Eq.\ref{eq:negativity_single}, and expand it for small $m$, 
\begin{equation}
E_N=\log\left\{     1-4 \left[    \cosh(\beta g) e^{-4\beta }  - \sinh(\beta g) e^{4\beta} \right]    \left\{  \frac{16\cosh(4\beta )}{1+4\cosh(4\beta +\cosh(8\beta ))}   +\frac{1728\beta^2 \left[  1+6\cosh(4\beta ) +\cosh(8\beta )\right]m^2 }{\left[   1+4\cosh(4\beta +\cosh(8\beta ))        \right]^2}   \right\}  \right\}.
\end{equation}
There the negativity $E_N$ is manifestly singular at $T_c$ due to the singularity from $m$.

\subsection{2c. Quantum Spherical Model }
Consider the Hamiltonian for a quantum spherical model: $H=\frac{1}{2} g \sum_{i=1}^N p_i^2 -\frac{1}{2N} \sum_{i,j=1}^N x_ix_j  + \mu \left[ \sum_{i=1}^N x_i^2 -\frac{N}{4} \right] $ where $[x_i,p_j] = i \delta_{ij}$. $\mu$ is chosen so that $\left< \sum_{i=1}^N x_i^2 \right> =\frac{N}{4}$ where the expectation value is taken with respect to the thermal density matrix. Define $x_k=\frac{1}{\sqrt{N}}  \sum_{j}e^{ikj} x_j$, $p_k=\frac{1}{\sqrt{N}}  \sum_{j}e^{ikj} p_j$ and introduce $a_k,a_k^{\dagger}$: $p_k=-i\sqrt{  \frac{\omega_{k}}{2g}} \left(   a_k - a_{-k} ^{\dagger} \right) $,$x_k=\sqrt{  \frac{g}{2\omega_{k}}} \left(   a_k + a_{-k} ^{\dagger} \right)$
the Hamiltonian can be diagonalized:
\begin{equation}\label{eq:spherical_diag_H}
H=\sum_{k} \omega_k  \left(a_k^{\dagger} a_k  +\frac{1}{2}\right)-\frac{\mu}{4}N,
\end{equation}
where the single particle energy $\omega_k$ is 
\begin{equation}
\omega_k=\begin{cases}
\omega_0=\sqrt{2g (\mu-\frac{1}{2})} \quad \text{for} \quad k=0\\
\omega_1=\sqrt{2g \mu}  \quad ~~~~~~~~ \text{for} \quad k\neq 0,
\end{cases}
\end{equation}
Note that in order to have a stable theory , $\mu\geq \frac{1}{2}$. From Eq.\ref{eq:spherical_diag_H}, the free energy density $f$ can be calculated:
\begin{equation}
f=\frac{1}{N\beta} \log\left[ 2\sinh (\frac{1}{2}\beta \omega_0)   \right] +\frac{N-1}{N\beta} \log\left[ 2\sinh (\frac{1}{2}\beta \omega_1)   \right] 
-\mu/4.
\end{equation}
$\mu$ is determined from $\left< \sum_{i=1}^N x_i^2 \right> =\frac{N}{4}$, which is equivalent to $\frac{\partial f}{\partial \mu}=0$:
\begin{equation}
\frac{1}{2N} \sqrt{\frac{g}{2(\mu-\frac{1}{2})}}  \coth\left( \frac{1}{2} \beta \sqrt {2g(\mu-\frac{1}{2})}\right) +\frac{N-1}{2N} \sqrt{\frac{g}{2\mu}}  \coth\left( \frac{1}{2} \beta \sqrt {2g\mu}\right) = \frac{1}{4}.
\end{equation}
In the thermodynamic limit $N\to \infty$, $\mu$ is a singular function of $\beta,g$. For $2\sqrt{g} \coth(\frac{1}{2}\beta \sqrt{g})>1$, the system is in a disordered phase, with $\mu$ determined from 

\begin{equation}\label{eq:spherical_mu}
\sqrt{\frac{g}{2\mu}} \coth(\frac{1}{2}\beta \sqrt{2g\mu})=\frac{1}{2},
\end{equation}
while the condition $2\sqrt{g} \coth(\frac{1}{2}\beta \sqrt{g})<1$ gives the ordered phase, and $\mu$ is pinned to $\frac{1}{2}$.
Here we brief describe the covariance matrix formalism for calculating the negativity of a Gaussian state $\rho$ for $N$ degrees of freedom. First we calculate the covariance matrix in displacements $(\gamma_x)_{ij}=\expval{\{ x_i-\overline{x}_i ,x_j-\overline{x}_j  \}} $ and the covariance matrix in momenta $(\gamma_p)_{ij}=\expval{\{ p_i-\overline{p}_i ,p_j-\overline{p}_j\}} $, where $\overline{x}_i=\tr{\rho x_i}$, $\overline{p}_i=\tr{\rho p_i}$, and $\{ A,B\}=AB+BA$ is the anticommutator. Define the subsystem $A$ composed by degrees of freedom for site $i=1,2,\cdots, N_A$ and the complement $B$ composed by the rest of sites, we calculate $\tilde{\gamma}=\gamma_xR\gamma_p R$, where $R$ is diagonal matrix with $1$ for the first $N_A$ diagonal entries and $-1$ for the rest of the diagonal entries. By diagonalizing  $\tilde{\gamma}$, we obtain its eigenvalues $\{\nu_i \vert i=1,2,\cdots, N \}$, from which the negativity $E_N$ can be calculated 

\begin{equation}
E_N(\rho)=\sum_{i=1}^{N}\text{max}\{0,-\log   \nu_i\}.
\end{equation}
For the thermal state of the spherical model, a straightforward calculation shows that
\begin{equation}
\begin{aligned}\label{eq:correlation}
&\left(\gamma_x \right)_{ij}=2\expval{x_i x_j} =m_x +\delta_{ij} d_x\\
&\left(\gamma_x \right)_{ij}=2\expval{p_i p_j} =m_p +\delta_{ij} d_p,
\end{aligned}
\end{equation}
with
\begin{equation}
\begin{aligned}
&m_x\equiv \frac{1}{N} \left[ \sqrt{\frac{g}{2\mu-1}}  \coth\left(\frac{1}{2} \beta \sqrt{(2\mu-1)g} \right)    -\sqrt{\frac{g}{2\mu}}  \coth\left( \frac{1}{2}\beta \sqrt{2\mu g}  \right) \right] \\
&d_x\equiv \sqrt{\frac{g}{2\mu}}  \coth\left( \frac{1}{2}\beta \sqrt{2\mu g}  \right)\\
&m_p\equiv\frac{1}{N} \left[ \sqrt{\frac{2\mu-1}{g}}  \coth\left(\frac{1}{2} \beta \sqrt{(2\mu-1)g} \right)    -\sqrt{\frac{2\mu}{g}}  \coth\left( \frac{1}{2}\beta \sqrt{2\mu g}  \right) \right] \\
&d_p\equiv \sqrt{\frac{2\mu}{g}}  \coth\left( \frac{1}{2}\beta \sqrt{2\mu g}  \right).
\end{aligned}
\end{equation}
Thus we have 
\begin{equation}
\tilde{\gamma}=\gamma_xR\gamma_p R=d_xd_p  \mathds{1}_N + m_xd_pJ_N + m_pd_x 
\begin{pmatrix}
J_{N/2} & -J_{N/2}\\
-J_{N/2}  & J_{N/2},
\end{pmatrix} 
\end{equation}
where we define $J_N$ as an $N\cross N$ all-ones matrix. All three matrices on the R.H.S. commute with each other so they can be diagonalized with the same set of eigenvectors. Since both the second and the third matrix are rank-1 matrix, it is easy to calculate the eigenvalues. Finally, the eigenvalues of $\tilde{\gamma}$ are
\begin{equation}
\nu_k= \begin{cases}
d_xd_p = \left[  \coth\left( \frac{1}{2}\beta \sqrt{2\mu g }   \right)  \right]^2 \quad \text{for} \quad  k=1,2, \cdots N-2\\
d_xd_p+ Nm_xd_p =\sqrt{\frac{2\mu}{2\mu-1}}  \coth\left(\frac{1}{2} \beta \sqrt{ (2\mu-1)g}  \right)  \coth\left( \frac{1}{2}\beta \sqrt{2\mu g }   \right)     \quad \text{for} \quad  k=N-1  \\
d_xd_p+ Nm_pd_x=\sqrt{\frac{2\mu-1}{2\mu}}  \coth\left(\frac{1}{2} \beta \sqrt{ (2\mu-1)g}  \right)  \coth\left( \frac{1}{2}\beta \sqrt{2\mu g }   \right)       \quad \text{for} \quad k=N 
\end{cases}
\end{equation}
One can check that $\nu_k>1 $ for $k=1,2,\cdots, N-1$ for all values of parameters in the model, and only $\nu_{N}$ can be less than $1$ to contribute to the entanglement negativity: 
\begin{equation}
\boxed{
	E_N=\text{Max}\{ 0, -\log \nu  \} }
\end{equation}
where 
\begin{equation}
\nu\equiv \nu_N=\sqrt{\frac{2\mu -1}{2\mu}}  \coth\left[ \frac{1}{2} \beta \sqrt{\left(g(2\mu-1)\right)}   \right] \coth\left[ \frac{1}{2} \beta\sqrt{2g\mu}  \right].
\end{equation}
By using Eq.\ref{eq:spherical_mu} in the disordered phase, and $\mu=\frac{1}{2}$ in the ordered phase, $\nu$ can be further simplified:
\begin{equation}
\nu=\begin{cases}
\frac{2}{\beta\sqrt{g}}   \coth(\frac{1}{2}\beta \sqrt{g}) \quad \quad \quad \quad \quad \quad \quad   \text{for ordered phase} \\
\frac{1}{2} \sqrt{\frac{2\mu-1}{g}} \coth(\frac{1}{2}\beta \sqrt{(2\mu-1)g})  \quad \text{for disordered phase}.
\end{cases}
\end{equation}
To study the singularity of $E_N$ at the critical point, we calculate the first derivative of $E_N$ with respect to $g$ to observe its discontinuity at a critical point:

\begin{equation}
\begin{split}
&\eval{\frac{ \partial E_N}{\partial g}}_{g_c^{+}}= \frac{1}{g_c} +\frac{\beta_c^2}{12} \left(1 - \frac{8}{4 + \beta_c - 4 \beta g_c} \right)\\
&\eval{\frac{  \partial E_N}{\partial g}}_{g_c^{-}}=\frac{4 + \beta_c - 4 \beta_c g_c}{8 g_c}	,
\end{split}
\end{equation}

\section{3. Entanglement of Formation in a Infinite-Range Commuting Projector Hamiltonian}
To begin with, we recall the definition of the entanglement of of formation: a density matrix $\rho$ acting on a bipartite Hilbert space $\mathcal{H}=\mathcal{H}_A\otimes \mathcal{H}_B$ can be decomposed as a convex sum of pure states
\begin{equation}
\rho=\sum_k P_k\ket{k}\bra{k},
\end{equation}
and for each $\ket{k}$, we can calculate the reduced density matrix on $A$: $\rho_k^A =\Tr_{B}  \ket{k}  \bra{k}$, from which the entanglement entropy $S_A(\ket{k})$ is obtained: $S_A(\ket{k})= -\Tr_A \rho_k^A \log \rho_k^A$. The entanglement of formation $E_F(A,B)$ is defined as 
\begin{equation}
E_F(A,B)=\text{min} \sum_k P_k S_A(\ket{k}),
\end{equation}
where minimization over all possible pure state decomposition is taken. Here we provide a model, where the entanglement of formation can be calculated analytically by showing its upper and lower bound coincide in the thermodynamic limit. Consider a one-dimensional lattice of size $L$ where each lattice site has two qubits, the model Hamiltonian is 
\begin{equation}
H=-\frac{1}{2L}  \left( \sum_{i=1}^L  Z_{i1}   Z_{i2}    \right)^2 -g \sum_{i=1}^L  X_{i1}   X_{i2} .
\end{equation}	
The density matrix at inverse temperature $\beta$ is $\rho=\frac{1}{Z}e^{-\beta H}$ with $Z=\Tr e^{-\beta H}$. We make an entanglement cut across one of the sites (say $s$-th site) such that the two spins on $s$-th site are not in the same subsystem. In the following calculation, $A$ comprises all the lattice sites with site index $i<s$ and the spin labelled by $1$ on $s$-th site while $B$ comprises all the lattice sites with site index $i>s$ and the spin labelled by $2$ on $s$-th site. For such a bipartition scheme, we prove that the entanglement of formation $E_F$ between $A$ and $B$ is exactly that from a mean-field density matrix for just two spins, where a closed form expression for $E_F$ is available. Our strategy is to find an upper bound and a lower bound on $E_F$ that happen to match each other. \\

\noindent\textbf{\textit{Upper Bound}}\\
Entanglement of formation $E_F$ requires a minimization scheme over all possible pure state decompositions. By considering a particular way of decomposition, we thus give an upper bound for $E_F$. First  we perform the Hubbard-Stratonovich transformation for $\rho$:

\begin{equation}\label{eq:hs_entanglement}
\rho =\frac{1}{Z}e^{-\beta H} =\frac{1}{Z}\sqrt{\frac{  \beta  L}{2 \pi   }} \int dm e^{ -\frac{1}{2}\beta L m^2     -     \beta \sum_{i=1 }^L H_i(m)},
\end{equation}
where a local Hamiltonian $H_i(m)$ for $i$-site of two spins is defined as :
\begin{equation}
H_i(m)= - m  Z_{i1}   Z_{i2}     -gX_{i1}   X_{i2}.
\end{equation}
Each $e^{-\beta H_i(m)}$ can be decomposed: $e^{-\beta H_i(m)}=\sum_{k_i} w^i_{k_i} (m)\ket{k_i(m)} \bra{k_i(m)}  $. As a result,
\begin{equation}
\rho=  \sum_{\{k_i\}}   \int dm   \frac{1}{Z} \sqrt{  \frac{\beta L}{2\pi}}   e^{-\frac{1}{2}\beta L m^2}   \left(\prod_{i} w^i_{k_i}(m) \right)   \ket{k_1,\cdots,k_L}\bra{k_1,\cdots,k_L} 
\end{equation}
The entanglement entropy between $A$ and $B$ in   $\ket{k_1,\cdots,k_L}\bra{k_i,\cdots,k_L} $ is given by the entanglement entropy between just two spins at site $s$ due to the product state structure for different sites. Therefore, 

\begin{equation}
E_F(A,B) \leq  \min_{\{k_i\}}  \sum_{\{k_i\}}   \int dm    \frac{1}{Z} \sqrt{  \frac{\beta L}{2\pi}}   e^{-\frac{1}{2}\beta L m^2}   \left(\prod_{i} w^i_{k_i}(m) \right)  S_{s1}(\ket{k_s(m)}),
\end{equation}
where $ S_{s1}(\ket{k_s(m)})$ is the entanglement entropy between spins at $s_1$ and $s_2$ in the state $\ket{k_s(m)}$, and the minimum is taken among all possible pure state decomposition of $e^{-\beta H_i(m)}$. Since $S_{s1}(\ket{k_s})$ is independent of how we decompose $e^{-\beta H_i}$ for $i\neq s$. The summation over $k_i ~\forall i\neq s$ can be performed on $w^i_{k_i}$:
\begin{equation}
\sum_{\{k_i  \vert i\neq s \}}  \prod_{i\neq s} w^i_{k_i} = \left(\Tr_i e^{-\beta H_i(m)}   \right)^{L-1}=e^{-\beta (L-1)f(m)},
\end{equation}
where $f(m)$ is a mean-field free energy density.
Consequently,
\begin{equation}
E_F (A,B)\leq \min_{k_s}   \frac{  \int dm e^{-\beta Lf(m)} \sum_{k_s}  \frac{1}{Z_s } w^s_{k_s}(m)  S_{s1}(\ket{k_s(m)}) } {\int dm e^{-\beta Lf(m)}},
\end{equation} 
with $Z_s\equiv \Tr_s e^{-\beta H_s(m)}$. In $L\to \infty $ limit, the argument inside the summation over $k_s$ is dominated only by saddle points, and thus 

\begin{equation}\label{eq:eof_result}
E_F(A,B)\leq  \min \sum_{k_s}\frac{1}{Z_s} w^s_{k_s}(m^*)  S_{s1}(\ket{k_s(m^*)}), 
\end{equation}
where $m^*$ is a saddle point obtained by minimizing $f(m)$. Define the mean field density matrix on a single site of two spins: 
\begin{equation}
\rho_s(m^*)= \frac{1}{Z_s} e^{-\beta H_s(m^*)},
\end{equation}
we show 
\begin{equation}\label{eq:upper}
E_F(A,B)\leq  E_F(s1,s2),
\end{equation}
i.e., the entanglement of formation between $A$ and $B$ is upper bounded by the entanglement of formation between two spins in the mean field density matrix.\\\\
\noindent\textbf{\textit{Lower Bound}}\\
As a bona fide entanglement measure, entanglement of formation is non-increasing under a partial trace. This implies that $E_F(a,b)\leq  E_F(A,B)$, where $a$ and $b$ denote a subsystem in $A$ and $B$ respectively. Here we choose two spins at the sites $s$ as $a$ and $b$. A calculation shows that the reduced density matrix at site $s$ is 
\begin{equation}
\rho_s=\frac{1}{Z}  \Tr_{i\neq  s} e^{-\beta  H} = \frac{\int dm   e^{-\beta L f(m)}    \frac{1}{Z_s}   e^{-\beta H_s(m)}  }{   \int dm e^{-\beta f(m)}   }=     \frac{\int dme^{-\beta Lf(m)}   \rho_s(m) }{   \int dm e^{-\beta f(m)}   } 
\end{equation} 
where $f(m)=-\frac{1}{\beta }    \log Z_s=-\frac{1}{\beta }   \log \Tr_s e^{-\beta H_s(m)}$ being the free energy density. In $L\to \infty $ limit, $\rho_s$ is exactly given by $\rho_s(m^*)$ where the saddle point $m^*$ is the location of the global minimum of $f(m)$. One way to see this is to expand $\rho_s$ in a complete operator basis on site $s$, and show that expectation value of any operator on site $s$ is precisely given by $\rho_s(m^*)$. This calculation shows that 
\begin{equation}\label{eq:lower}
E_F(s1,s2)\leq  E_F(A,B). 
\end{equation}
By combining Eq.\ref{eq:upper} and Eq.\ref{eq:lower}, one finds that the bi-partite entanglement of formation between $A$ and $B$ is exactly that between two spins in the mean field density matrix which can be calculated analytically using the result of Ref.\cite{wooters1997}.
 
\end{document}